\documentclass[12pt]{article}
\usepackage{latexsym,graphicx,multirow}
\usepackage{float}
\usepackage{amssymb}
\usepackage{amsmath}
\usepackage{amscd}
\usepackage{amsthm}
\usepackage[left=2cm,top=2.5cm,right=2.5cm,bottom=1.5cm]{geometry}
\usepackage{hyperref}
\usepackage{epstopdf}
\usepackage{cite}
\begin{document}
	\begin{center}
	\large{\bf{Barrow holographic dark energy with Hubble horizon as IR cutoff}} \\
	\vspace{10mm}
	\normalsize{ Shikha Srivastava$^1$, Umesh Kumar Sharma$^2$ }\\
	\vspace{5mm}
	
	{$^{1,2}$ Department of Mathematics, Institute of Applied Sciences and 	Humanities, GLA University\\
			Mathura-281406, Uttar Pradesh, India}\\
	 
	\vspace{2mm}
		$^1$E-mail: shikha.azm06@gmail.com\\
         $^2$E-mail:  sharma.umesh@gla.ac.in\\
         
			\vspace{2mm}
		\vspace{5mm}
		\vspace{10mm}
	
		
\end{center}
\begin{abstract}
	
 In this work, we propose a non-interacting model of Barrow holographic dark energy (BHDE) using Barrow entropy in a spatially flat FLRW Universe considering the IR cutoff as the Hubble horizon. We study the evolutionary history of important cosmological parameters, in particular,  EoS $(\omega_{B})$,  deceleration parameter and, the BHDE and matter density parameter and also observe satisfactory behaviours in the BHDE the model. In addition, to describe the accelerated expansion of the Universe the correspondence of the BHDE model with the quintessence scalar field has been reconstructed.\\
	
\end{abstract}


\section{Introduction}
According to a sequence of past and latest observational data  \cite{refs1,refs2,refs3,refs4,refs5,refs6}, the dark sector of our Universe is filled  with two dark fluids, namely the dark energy (DE)
and dark matter (DM). The former fluid drives the current accelerating stage of the Universe while the later is responsible for the structure formation of the Universe. It is also observed from the observational data that about ninety-six per cent of the total energy density of the Universe is coming from this combined dark component, where the contribution of DE is around sixty-eight per cent of the total energy allocation of the Universe while the DM contributes approximately twenty-eight per cent of the total energy allocation of the Universe. Although, the origin, evolution and the characteristics of these dark sector of the DE are not distinctly recognized yet \cite{refs7,refs8,refs9,refs10,refs11,refs12}. However, the nature of DM appears to be partially known by indirect gravitational effects, although, the DE has endured being exceptionally mysterious. As a consequence, in the last couple of years, various cosmological models have been proposed and
explored.  The non-interacting models of the Universe are considered to be the simplest model leading to two independent evolution of these dark components where DE and DM are conserved separately. In a more general form of available cosmological models  interaction between DE and DM are permitted.  \cite{ref6a}.\\

The quantitative description of dark energy can suitably be substituted by holographic dark energy (HDE) \cite{ref7,ref8},  deriving from  holographic principle (HP) \cite{ref1,ref2,ref3,ref4,ref5}. 
Notably one can recieve the consequence of vacume energy of holographic source at  astrophysical scales structure DE \cite{ref7,ref8} due to the relationship between the  longest length of  $QFT$ (quantum field theory) with its UV cutoff \cite{ref6}. Interestly both the simple \cite{ref9,ref10,ref11,ref12,ref13,ref14,ref15,ref16,ref17,ref18}  and  extended \cite{ref19,ref21,ref23,ref24,ref25,ref26,ref27,ref28,ref29,ref30,ref31,ref32,ref33,ref34,ref35,ref36,ref37,ref38,ref39,ref40,ref41,ref42,ref43,ref44} versions of HDE represents to lead to a fascinating cosmological behaviour and  they also constrained with observations \cite{ref45,ref46,ref47,ref48,ref49,ref50,ref51,ref52,ref53}. The Universe horizon  entropy remains proportional to its area, which becomes the most pevetel step in the application of HP at cosmological context, similar to  a black hole BH (Bekenstein-Hawking) entropy. Although,  for the black-hole structure, recently Barrow propoesd that
the QG (quantum-gravitational) impacts can lead to  introduce fractal, intricate behaviours.
He emphasized that this complex structure gives finite volume but with the finite (or infinite) area and  hence a deformed expression of black-hole
entropy \cite{ref54}

\begin{eqnarray}
\label{eq1}
S_{B}=\left(\frac{A}{A_{0}}\right)^{\frac{2+\Delta}{2}},
\end{eqnarray}

where, $A_0$  and $A$ are  Planck area and standard horizon area, respectively. A new exponent $\Delta$, contributing to the quantum-gravitational deformation was in $\Delta$ = 1 corresponding to
its the most fractal and intricate structure and corresponds to the usual BH entropy when $\Delta$ = 0.  It is important to mention that  the standard ``quantum-corrected''
entropy with  corrections logarithmic \cite{ref55,ref56} is  different from the above quantum-gravitationally corrected entropy.  However, it resembles non-extensive Tsallis entropy \cite{ref57,ref58,ref59},  but the physical principles and involved foundations are totally different.\\

Recently, Saridakis \cite{ref60}, constructed the BHDE, by using the usual HP, however applying the Barrow entropy instead of the BH entropy. Also,  for the limiting case as $\Delta$ = 0, the BHDE possesses standard HDE, although The BHDE, in general, is a new scenario with nice cosmological behavior and richer structure.  While standard HDE is given by the inequality $\rho_{B} L^{4} \leq S$, where $L$ denotes  horizon length, and under the imposition $ S \propto A \propto L^{2}$ \cite{ref7}, the Eq. (\ref{eq1}) will give
\begin{equation}
\label{eq2}
\rho_{B}=C L^{\Delta-2},
\end{equation}
where $C$  is a parameter having  $[L]^{-2-\Delta}$ dimensions \cite{ref60}.  As expected, in the case $\Delta=0$, the above expression gives the standard HDE  $\rho_{B}=3 c^{2} M_{p}^{2} L^{-2}$, where $C=$ $3 c^{2} M_{p}^{2}$  with $c^{2}$ the model parameter and  $M_{p}$ is the Planck mass. Depending on the parameter $\Delta$,  the BHDE will deviate by the standard one,  which may lead to distinct cosmological behaviour. If we  take into consideration the IR cut off $L$ as the Hubble horizon $(H^{-1})$, then  the energy density of BHDE is obtained as 
\begin{equation}
\label{eq3}
\rho_{B}=C H^{2-\Delta}.
\end{equation}\\
Saridakis \cite{ref61}, using Barrow entropy presented a modified cosmological scenario besides the Bekenstein-Hawking one. For the evolution of the effective DE density parameter, the analytical expression was obtained and shown the DM to DE  era of the Universe.  Using the Barrow entropy on the horizon in place of the standard Bekenstein-Hawking one,  the potency of the generalized second law of thermodynamics has also been examined \cite{ref61a}.  Mamon et al. \cite{ref61b} studied interacting BHDE model and also the validity of the generalized second law by assuming dynamical apparent horizon as the thermodynamic boundary. More recently,
Anagnostopoulos et al. \cite{ref62} have shown that the BHDE
is in concurrence with observational information, 
and it can serve as a decent contender for the depiction of DE.  The authors examined the Tsallis nonextensive form of the logarithmically amended Barrow entropy \cite{ref62a}. Barrow's new idea of entropy 
have some significant examinations \cite{ref62b,ref62c}. Now, we discuss the similarity and difference with other work in literature. Recently, an interacting model of the BHDE has been proposed by using Barrow entropy.  In particular, the evolution of a spatially flat FLRW Universe filled
with BHDE and pressureless DM that interact with each other through a well-motivated
interaction term has been investigated by taking the  Hubble horizon as IR cutoff \cite{ref61b}. While in this work we focus the BHDE model without interaction in a flat FLRW Universe with Hubble horizon as IR cutoff.  We organize the present work in the following way. The cosmological parameters of BHDE model are discussed in  Sect. $2$.   In Sect. $3$, we investigate the stability of the BHDE model. The correspondence between the BHDE and quintessence scalar field model and also potentials for scalar field models are discussed in Sect. $4$. In Sect. $5$, we draw our conclusions.

\section {The Cosmological Model } 
The first Friedmann equation for a flat FRW Universe  which is composed by pressureless DM ($\rho_{m}$) and BHDE ($\rho_{B}$) is
\begin{equation}
\label{eq7}
H^{2}=\frac{1}{3 m_{p}^{2}}\left(\rho_{B}+\rho_{m}\right),
\end{equation}
expounding the dimensionless density parameter as $\Omega_{i}=\rho_{i} / \rho_{c},$ where $\rho_{c}=$ $3 m_{p}^{2} H^{2}$ is known as the critical energy density,  we can obtain
\begin{eqnarray}
\label{eq8}
\Omega_{B}=\frac{\rho_{B}}{3 m_{p}^{2} H^{2}}=\frac{C}{3 m_{p}^{2}} H^{- \Delta}, \quad \Omega_{m}=\frac{\rho_{m}}{3 m_{p}^{2} H^{2}}.
\end{eqnarray}

The conservation law corresponding to dust and BHDE are 
\begin{equation}
\label{eq10}
\dot{\rho}_{m}+3 H \rho_{m}=0,
\end{equation}
\begin{eqnarray}
\label{eq6}
\dot{\rho}_{B}+3 H \rho_{B}\left(1+\omega_{B}\right)=0.
\end{eqnarray}

Here $\omega_{B}=p_{B} / \rho_{B}$ is EoS parameter and $p_{B}$  is pressure of BHDE. Differentiating Eq. (4) with respect to time  and solving   Eqs. (6),  (7) and Eq. (5), we can obtain
\begin{equation}
\label{eq11}
\frac{\dot{H}}{H^{2}}=-\frac{3}{2}(1+\omega_{B} \Omega_{B}).
\end{equation}
Substituting  Eq. (3) in Eq. (7), we will find 
\begin{equation}
\label{eq12}
\frac{\dot{H}}{H^{2}}=-\frac{3(1+\omega_{B})}{2-\Delta},
\end{equation}

by the help of  Eq. (7) and Eq. (8), we get

\begin{equation}
\label{eq13}
\omega_{B}=-\frac{\Delta}{2-(2-\Delta) \Omega_{B}}.
\end{equation}
From Eq. (8), we can  find that
\begin{equation}
\label{eq15}
\Omega_{B}^{\prime}=\frac{d \Omega_{B}}{d(\ln a)}=(1- \Delta) \Omega_{B} \frac{\dot{H}}{H^{2}}.
\end{equation}
Associating  the above result with Eqs. (9) and (10), we obtains
\begin{equation}
\label{eq16}
\Omega_{B}^{\prime}=-3(1-\Delta) \Big(\frac{1-\Omega_{B}}{2-(2-\Delta) \Omega_{B}}\Big) \Omega_{B}.
\end{equation}


\begin{figure}
	\begin{center}
		\includegraphics[width=16cm,height=10cm, angle=0]{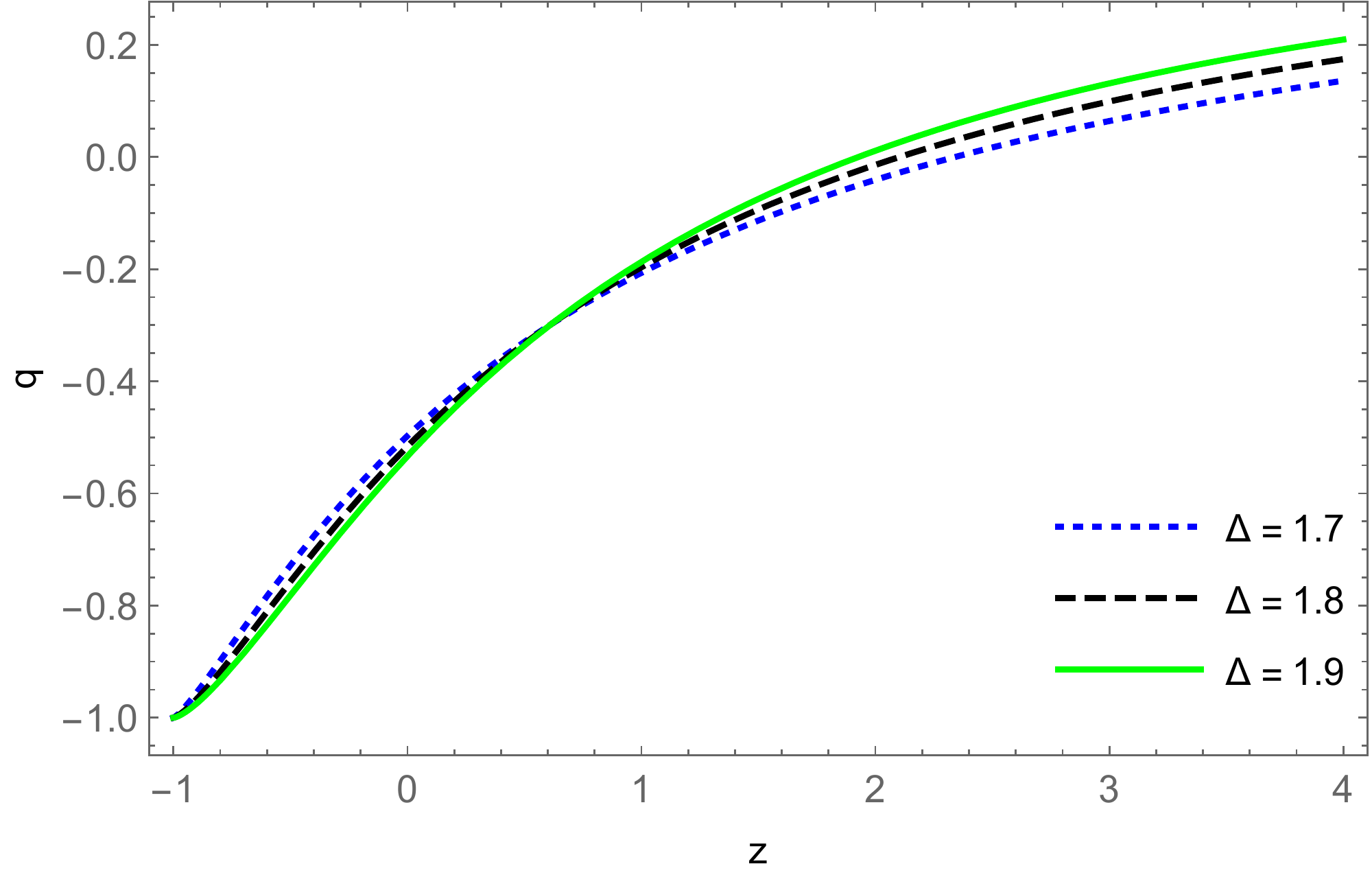}
		\caption{ Expansion of  deceleration parameter verses redshift $z$ distinct estimations of $\Delta$. Here, we take  $\Omega_{B0}=0.70$.}
	\end{center}
\end{figure}

As we know that deceleration parameter is given as
\begin{equation}
\label{eq18}
q=-\big[1+\frac{\dot{H}}{H^{2}}\big],
\end{equation}

putting Eq. (9)  in Eq. (13) and  Using Eq. (10), we get
\begin{equation}
\label{eq19}
q=\left[\frac{1-(1+ \Delta) \Omega_{B}}{2-(2-\Delta) \Omega_{B}}\right].
\end{equation}

The evolutionary behaviour of the deceleration parameter is plotted for the BHDE model versus redshift $z$ by finding its numerical solution using the initial values of $\Omega_{B}$ as $\Omega_{B0}$= 0.70. The value of deceleration parameter $q$ decides the nature of the Universe such as if $q<0$, it is accelerating and if  $q >0 $, it is in decelerating phase. It is proposed by various observations that the Universe is in an accelerated expansion phase and the value of deceleration parameter lies in the range $-1 \leq q < 0$. The evolutionary behaviour of the deceleration parameter versus redsfit $z$ is plotted in Fig. 1. We observe from this figure that the deceleration parameter of the BHDE model transits from an early decelerated phase to the current accelerated phase for the different values of the parameter $\Delta$. The evolutionary behaviour of the EoS parameter $\omega_{B}$ versus $z$	 for the BHDE model is plotted in Fig. 2 for the different values of the parameter $\Delta$.  
We can observe that the EoS parameter of the BHDE remains in quintessence era, and approaches to the cosmological constant ($\omega_{B} = -1$) at future for $\Delta$ =1.7, 1.8 and 1.9. It is significant that EoS parameter of the BHDE gives pleasant conduct and it tends to be quintessence-like for the various estimations of $\Delta$.\\
\begin{figure}
	\begin{center}
		\includegraphics[width=16cm,height=10cm, angle=0]{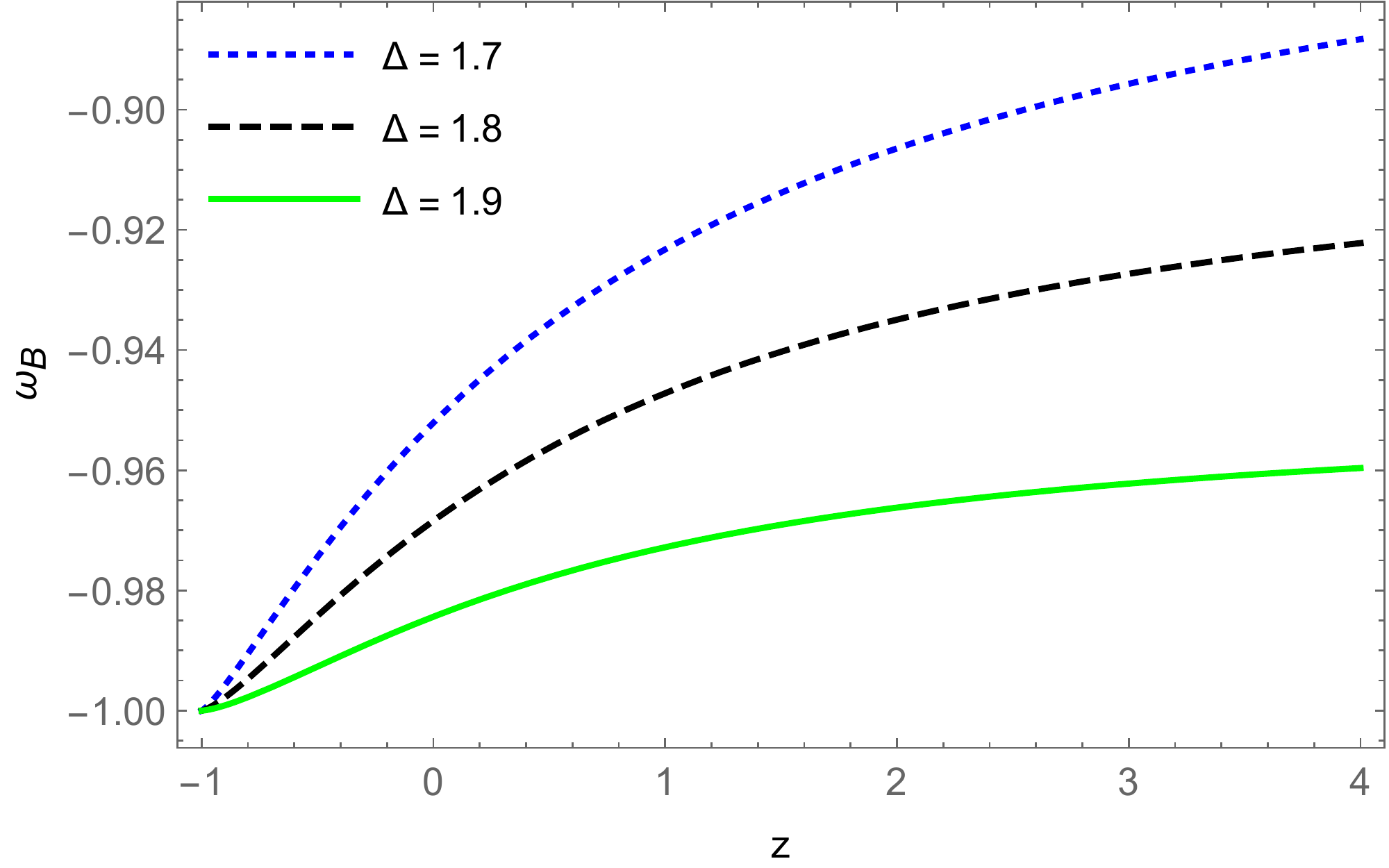}
		\caption{ The behaviour of the  EoS parameter of BHDE $\omega_{B}$   against $z$ for different $\Delta$. Here, we take  $\Omega_{B0}=0.70$. }
	\end{center}
\end{figure}

\begin{figure}
	(a)\includegraphics[width=8cm,height=8cm, angle=0]{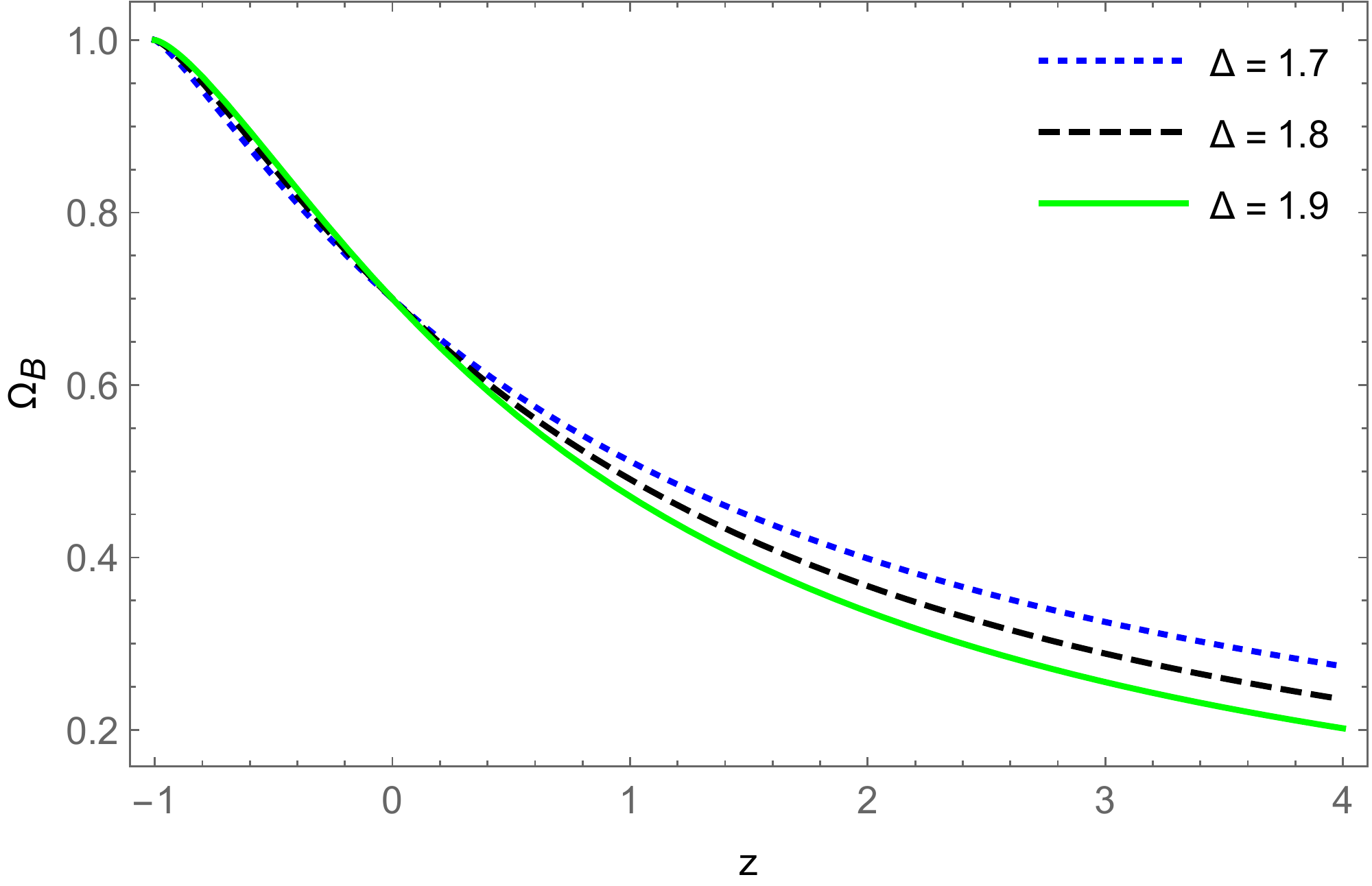}
	(b)\includegraphics[width=8cm,height=8cm, angle=0]{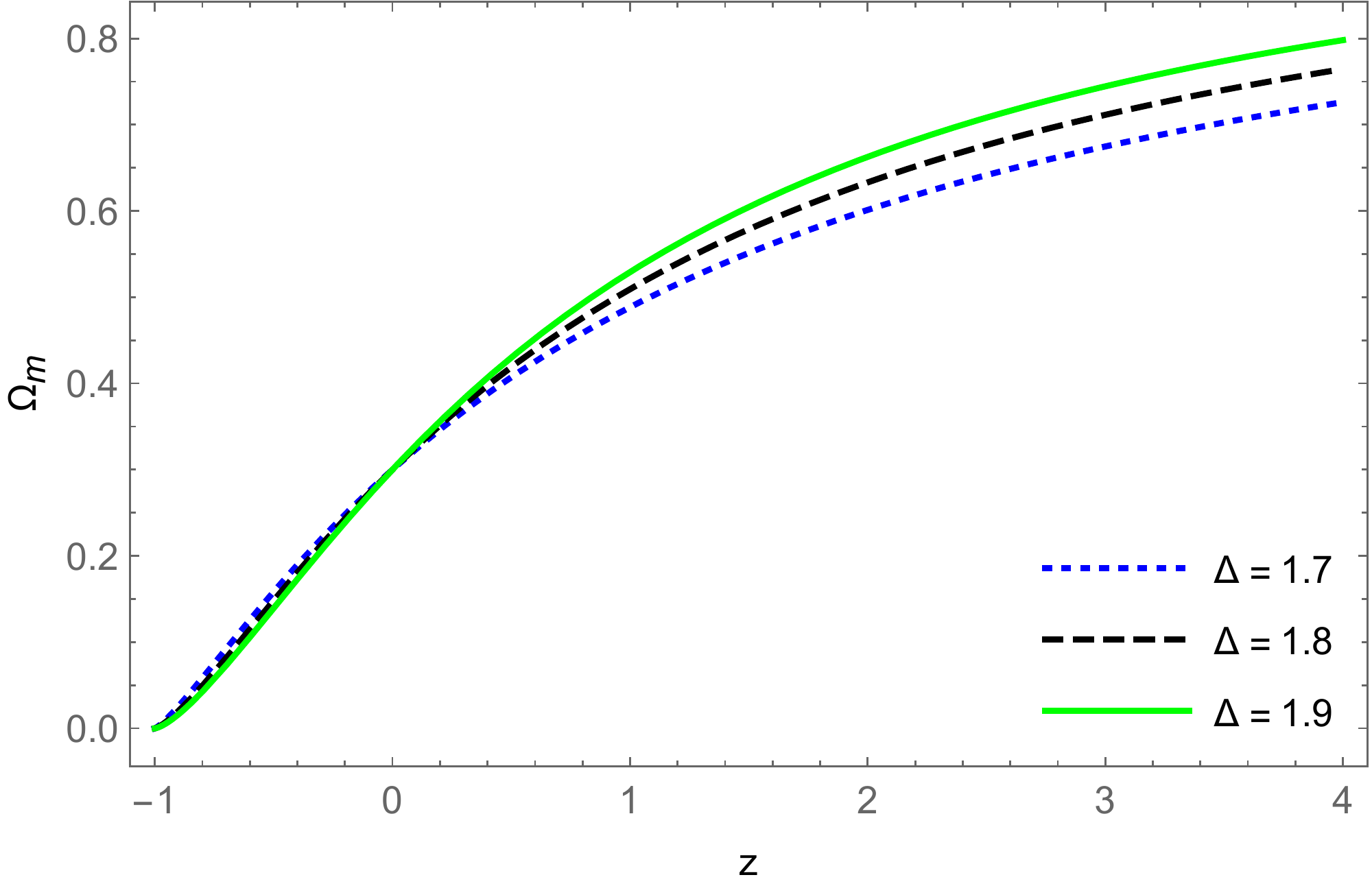}
	\caption{(a) Expansion  of  BHDE density parameter  $\Omega_{B}$ with $z$ redshift for distinct value of $\Delta$.Here, we take  $\Omega_{D0}=0.70$ (b) The evolution of  matter energy density parameter $\Omega_{m}$  with  $z$ (redshift)  for distinct value of $ \Delta$.  Here, we take  $\Omega_{B0}=0.70.$} 
\end{figure}
The  modified cosmological scenario has been constructed by applying the  FLT (first law of thermodynamics) and using the Barrow entropy \cite{ref61}. Moreover, Saridakis \cite{ref61}, proposed this cosmological modified scenario with varying parameter $\Delta$ by applying the non-extensive thermodynamics. In \cite{ref60,ref61,ref61a}, the researcers proposed that this cosmological modified scenario gives a description of both inflation and late-time acceleration of the Universe with  varying parameter from the FLT. In Fig. 3(a) and 3(b) we have plotted the energy density for the BHDE $\Omega_{B}$ and energy density of matter $\Omega_{m}$ as a
function of redshift.  The  thermal history of the Universe, in particular, the successive
sequence of matter and DE era can be observed from these figures.  In the papers \cite{ref60,ref61,ref61a}, it has also been  proposed by some of the researchers, relating to  the late times Universe evolution, a new
term that appear due to the  varying non-extensive exponent
constituted an effective DE sector. Futher, it has been shown that 	Universe shows usual thermal history, with the successive
sequence of matter and De epochs, and  the 
transition to acceleration is  in
agreement with the observed behavior.\\

\section{ Analysis of  BHDE Model for its Stability  }
Squared sound speed lies in a range $0\leq v^2_s\leq 1$ physically \cite{ref138b}. We take EoS parameter $\omega_{B} \neq -1$ to analyse the impact of the DE sound speed, because dark energy is of the form of cosmological constant consist of no perturbations so there is lack of sensitivity to the DE speed of sound squared. Effect of sound speed of dark energy on the CMB power spectrum  is detailed has analysed in \cite{ref138c} concluding that there is no impact of LISW (late-time integrated Sachs-Wolfe) with the decrement of $v^2_s$ from  $1$ to $0$ and when $ \omega_{B} >-1$  but when $ \omega_{B} <-1$ LISW gives positive impact. Consequently, we can see more clustering of (cold) dark energy with a decrement in $\omega_{B} > -1$ resulting in an increment of dark energy perturbations which is having the power to stop the potential to vanish and preserve it and thus introducing the LISW effect a small push \cite{ref138b}.\\

Stability Test for BHDE Model  against the perturbation has been done using squared speed of  sound $v^2_s =\frac{dp_T}{d\rho_{B}}$, \cite{ref138a}. Generally, we give the squared  speed of sound in the form:
\begin{eqnarray}
\label{eq18}
v^2_s =\frac{dp_B}{d\rho_{B}}
= \frac{2\Delta(1- \Delta) \Omega_{B}-2\Delta+\Delta \Omega_{B}}{[2-(2-\Delta) \Omega_{B}]^2}.
\end{eqnarray}


\begin{figure}[H]
	\begin{center}
		
		\includegraphics[width=16cm,height=8cm, angle=0]{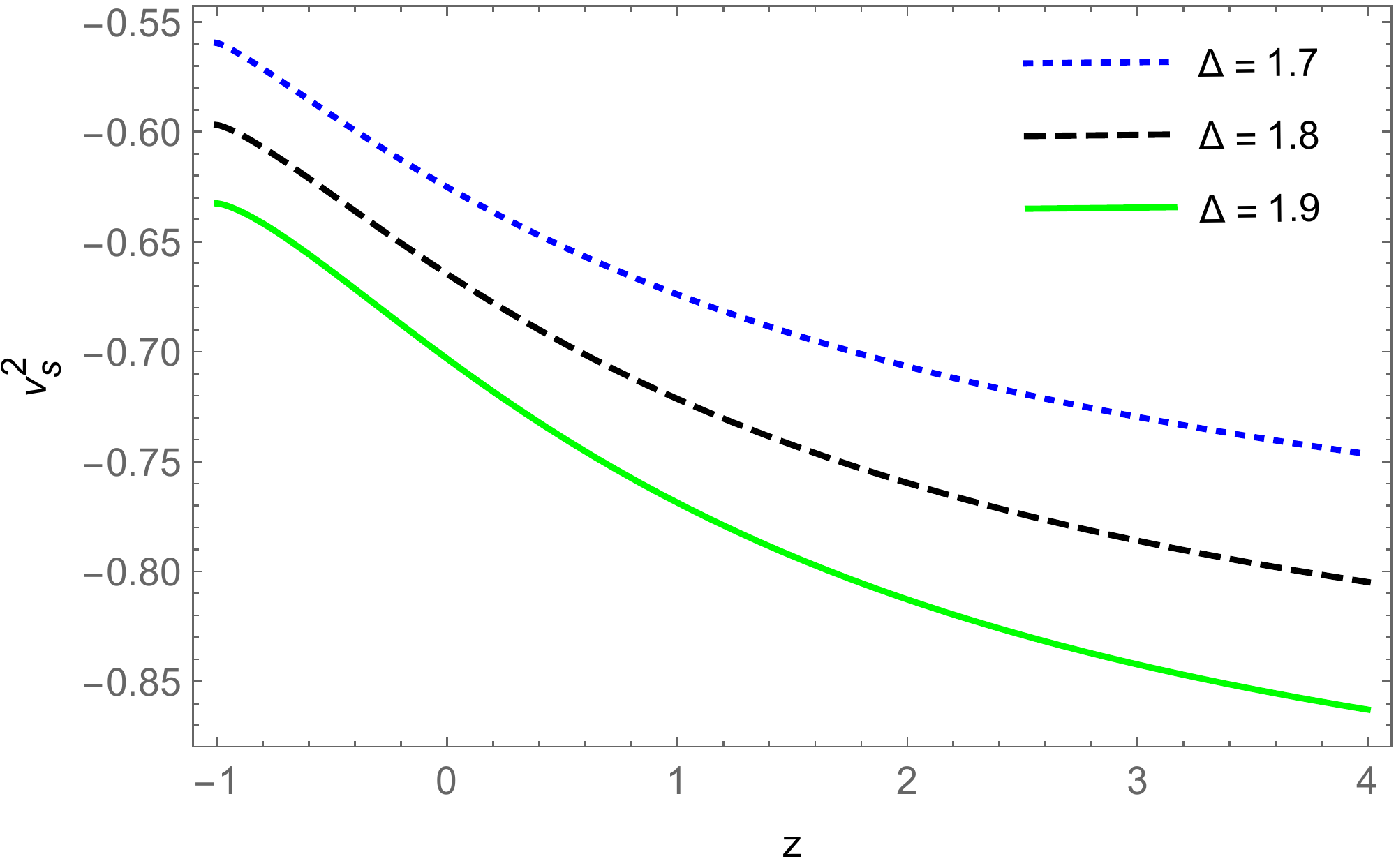}
		
		\caption {The behaviour of squared sound speed $v^2_s$ verses  redshift $z$ for distinct value of $\Delta$.  Here, we take  $\Omega_{B0}=0.70.$}
	\end{center}
\end{figure}

The squared speed of sound $v^2_s$ represented by Eq. (15), is plotted in  Fig. 4 as a function of redshift ($z$) for different values of the parameter $\Delta$,  and it can be observed by the figure that for $\Delta <2$, The BHDE model is unstable during the cosmic evolution.

\begin{figure}
	\begin{center}
		\includegraphics[width=10cm,height=12cm, angle=0]{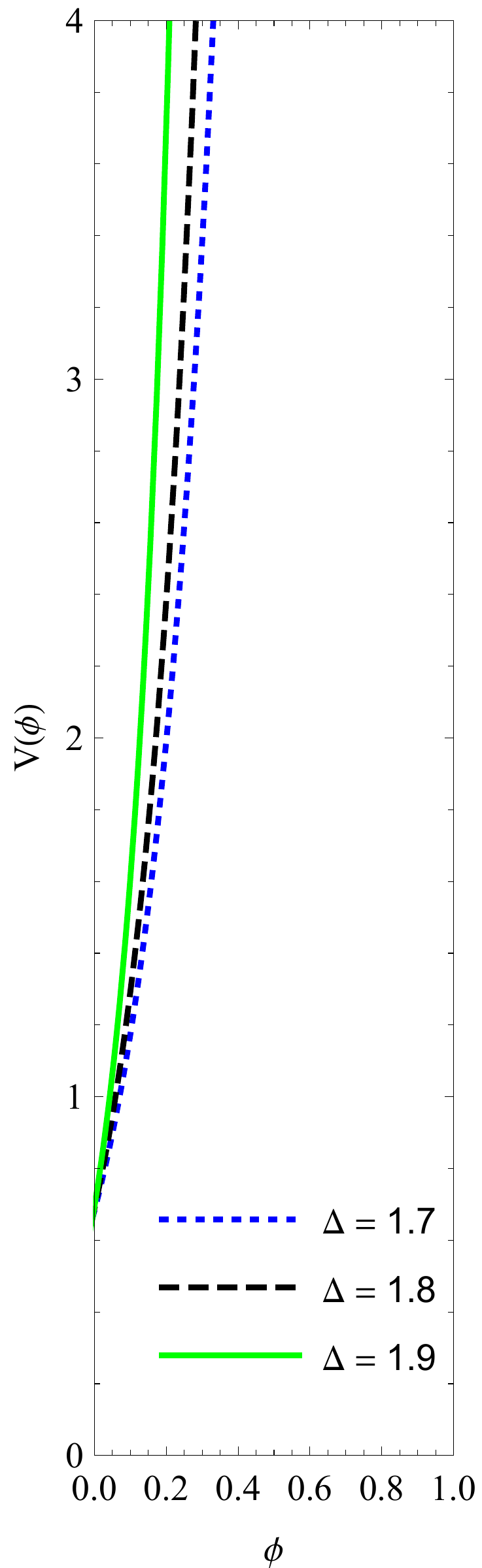}
		\caption{The  potential reconstruction for the Barrow holograhpic quintessence for different $\Delta$, 
			here $V(\phi)$  and $\phi$ are taken in unit of $\rho_{c 0}$ and  ${M_{p}}^2$ respectively and $\Omega_{B0}=0.70$ .}
	\end{center}
\end{figure}

\begin{figure}
	\includegraphics[width=16cm,height=8cm, angle=0]{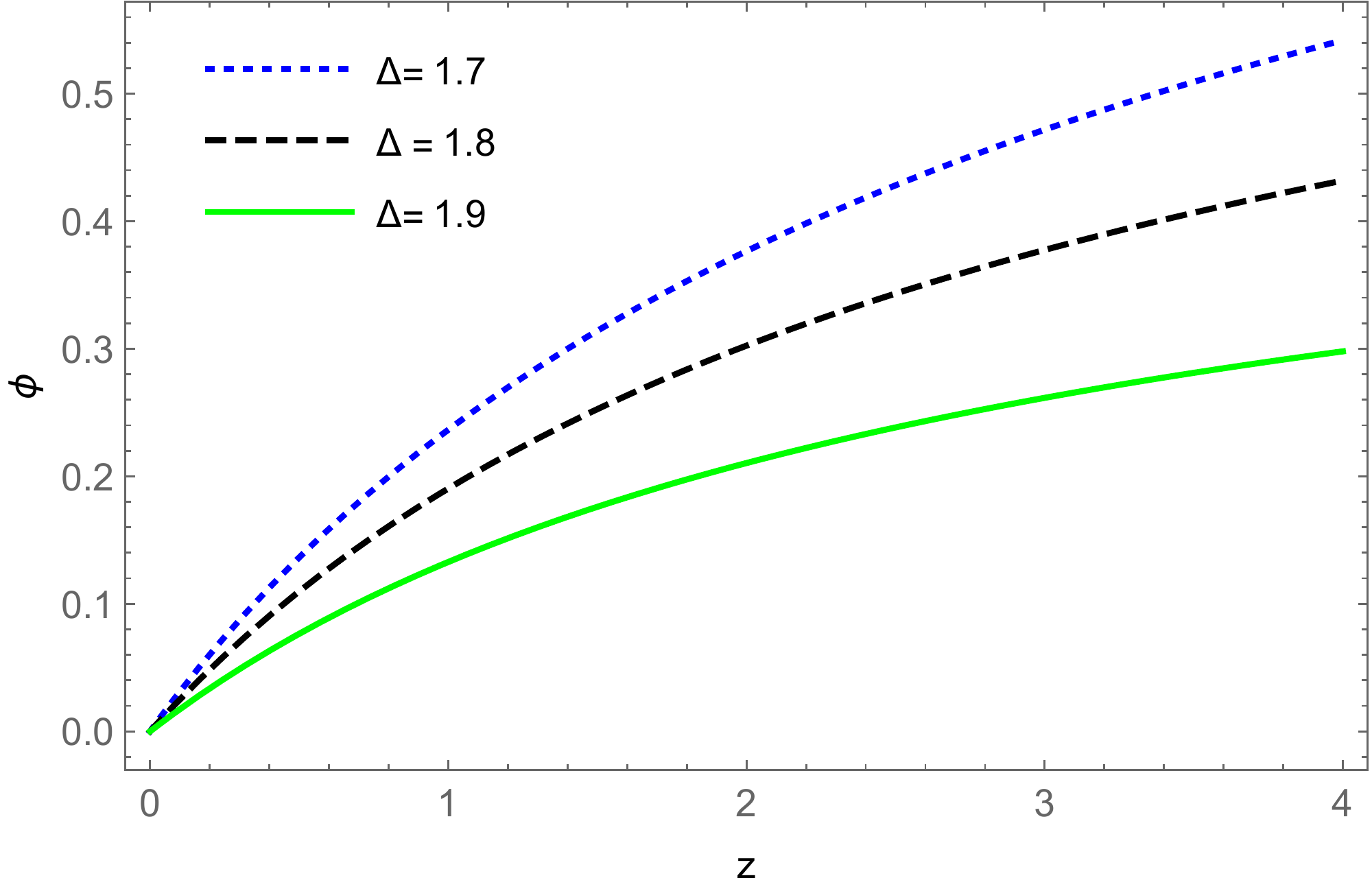}
	\caption{The evolution of the scalar-field $\phi(z)$ for quintessence of  Barrow holograhpic, here  $\phi$  is taken in unit of ${M_{p}}^2$ and $\Omega_{B0}=0.70$.}
\end{figure}

\section{ The correspondence of BHDE with   quintessence  field}

In view of this segment, we will deliberate the correspondence of the BHDE with  quintessence scalar field model. 
The potentials and dynamics for  quintessence scalar field
are also reconstructed.  By comparing energy densities of the scalar field models and the BHDE model given in Eq. (\ref{eq3}), we obtain the correspondence. Here we also compare the EoS parameter for scalar fields (quintessence ) with EoS of BHDE  model stated by Eq. (\ref{eq13}). Both the dark energy,  canonical and non-canonical are well described by using scalar fields. In the present article, we have chosen the quintessence field as canonical. As we know that  for quintessence  scalar field $\omega_{B} > -1$   \cite{ref116,ref116a,ref117,ref120,ref127}. 	\\

The pressure and energy density for quintessence scalar field \cite{ref139a} are respectively given below as:

\begin{eqnarray}
\label{eq19}
\quad p = \frac{\dot{\phi}^{2}}{2} - V(\phi),  \rho = \frac{\dot{\phi}^{2}}{2} + V(\phi),
\end{eqnarray}

 where $ V(\phi )$ stands for scalar field potential and $\phi$ stands for scalar field. 
	\begin{eqnarray}
	\label{eq20}
	V(\phi)=\frac{1-\omega_{B}}{2} \rho_{B},
	\end{eqnarray}
	\begin{eqnarray}\label{eq21}
	\dot{\phi}^{2}=\left(1+\omega_{B}\right) \rho_{B}.
	\end{eqnarray}
	Moreover, from the equation of a flat  $\mathrm{FLRW}$  $3 M_{p}^{2} H^{2}=\rho_{m}+\rho_{B}$, we can obtain
	\begin{eqnarray}\label{eq22}
	E(z)\equiv \frac{H(z)}{H_{0}}=\left(\frac{\Omega_{m 0}(1+z)^{3}}{\left(1-\Omega_{B}\right) }\right)^{1 / 2},
	\end{eqnarray}
	here $\Omega_{m 0}$ represents the  present fractional energy density of pressureless matter. By the help of  Eq. $(19)$ we can express Eqs. (17) and (18) respectively \cite{ref139a}
	\begin{eqnarray}
	\frac{V(\phi)}{\rho_{c 0}}=\frac{1}{2}\left(1-\omega_{B}\right) \Omega_{B} E^{2},
	\end{eqnarray}
	\begin{eqnarray}
	\frac{\dot{\phi}^{2}}{\rho_{c 0}}=\left(1+\omega_{B}\right) \Omega_{B} E^{2}.
	\end{eqnarray}
	Where $\rho_{c 0}=3 M_{P}^{2} H_{0}^{2}$ gives present  critical density of the Universe. For the  propose  correspondence between the BHDE and the quintessence scalar field, namely, in particular, we recognize $\rho$ with $p$. Futhermore, quintessence field takes over Barrow nature such as $E, \Omega_{B}$ and $\omega_{B}$ are represented by Eqs. (10), (12) and $(19)$. According to  \cite{ref139a}, we assume $\dot{\phi}>0$ in this paper. Differentiating  $\phi$ (scalar field)  with respect to  $z$ (redshift), given as
	\begin{eqnarray}
	\frac{\frac{d \phi}{d z}}{M_{p}}=\frac{\sqrt{3\left(1+\omega_{B}\right) \Omega_{B}}}{1+z}.
	\end{eqnarray}
	
	By  integrating Eq. (22),  we can without much of a stretch acquire the evolutionary structure of the field as :
	
	\begin{eqnarray}
	\phi(z)=\int_{0}^{z} \frac{d \phi}{dz} dz,
	\end{eqnarray}
	here, the field amplitude ($\phi(z)$), at the present epoch $(z=0)$ is constant to be zero such as $\phi(0)=0$.  In this manner, this can  be constituted as Barrow Holographic quintessence DE model and recreate the potential of the $\mathrm{BHDE}$.

	In Fig. ${5}, $ $V(\phi)$,  the reconstructed quintessence potential  is graphed, and  $\phi(z)$   which is also recreated by Eqs. (22) and $(23),$  plotted in Fig. $6$. Assuming present fractional energy density is  $\Omega_{\mathrm{B} 0}=0.70.$, chosen curves are portrayed for the different cases of $\Delta$. The dynamics of the scalar field explicitly observed from Figs. $5$ and $6$. Clearly, the scalar field $\phi$ moves down the potential with kinetic energy $\dot{\phi}^{2}$ slowly diminishing.


\section{Conclusions}
 In this paper, we have constructed the BHDE model without interaction which depends on Barrow
	entropy proposed by Barrow recently, it begins with the modification proposal of
	the black-hole structure because of some QG effects. We have studied the evolution of a spatially
	flat FLRW Universe composed of the BHDE and pressureless DM by considering the IR cutoff as Hubble horizon. We have explored the behaviour of cosmological parameters such as  DP (deceleration parameter), the  EoS parameter, the BHDE and matter-energy density parameter during the cosmic evolution. The correspondence between BHDE model and the quintessence scalar field models have also been done to clarify the late-time cosmic accelerated expansion of the Universe. We finish up our outcomes as follows
	: \\

	$\bullet$ It has been observed that the BHDE model exhibits a smooth transformation from an early deceleration era to present acceleration era of the Universe and in a good agreement with recent cosmic observations.\\

	$\bullet$ The behaviour of EoS parameter can be observed by varying the exponent  $\Delta$. The EoS parameter of the BHDE model varies in the quintessence region for different values of the parameter $\Delta$.\\

	$\bullet$ 	We extricated a straightforward  DE (Differential Equation) for the evolution of the DE density parameter, and we presented the behaviour of  BHDE and matter density parameter. As we appeared,
	the situation of the BHDE can portray the Cosmos usual history, with the arrangement of matter and DE era.  \\

	$\bullet$ We observe that for $\Delta <2$, the BHDE model is not stable for all values of $z$  during the cosmic evolution. It can be fixed by taking other IR cutoffs, probable interactions between the Universe sectors, different entropy corrections or even a mix of these scenarios.
	Indeed, this contemplation can likewise modify and increase predictions and behaviour of  BHDE. These are subjects concentrated in future to turn out to be all the more near the various properties of BHDE, and thus the cause of the DE.\\

	$\bullet$ We proposed  a correspondence between  BHDE model and
	quintessence scalar-field model. We have a look at that the BHDE with $\Delta <2$ may be described absolutely with the aid of the quintessence in a sure manner. The correspondence between  BHDE and quintessence has been hooked up, and the potential of the Barrow holographic	quintessence and the dynamics of the field have been recreated.	\\

\section{Acknowledgements}
The author S. Srivastava thankfully recognize the utilization of administrations and office gave by GLA University, Mathura, India to lead this research work. The author U. K. Sharma thanks the IUCAA, Pune, India for awarding the
visiting associateship.

\end{document}